\documentclass[article,aps,nofootinbib,showpacs,amsfonts,epsf]{revtex4}

\input{epsf.tex}

\newcommand{\E}{{\cal{E}}}

\newcommand{\s}{\sigma}

\renewcommand{\d}{{\rm d}}
\renewcommand{\a}{\alpha}

\newcommand{\be}{\begin{equation}}
\newcommand{\ee}{\end{equation}}
\newcommand{\bea}{\begin{eqnarray}}
\newcommand{\eea}{\end{eqnarray}}
\newcommand{\ba}{\begin{array}}
\newcommand{\ea}{\end{array}}
\def\J#1#2#3#4{{#1} {\bf #2}, #3 (#4)}
\def\PRD{Phys. Rev. D}
\def\PR{Phys. Rev.}
\def\PRL{Phys. Rev. Lett.}
\def\PTP{Prog. Theor. Phys.}
\def\APN{Ann. Phys. (NY)}

\def\TMP{Theor. Math. Phys.}

\def\JHEP{J. High Energy Phys.}

\def\CQG{Class. Quantum Grav.}

\def\GRG{Gen. Relativ. Grav.}
\def\PLA{Phys. Lett. A}

\begin{document}
\draft
\title{Schwarzschild black hole levitating in the hyperextreme Kerr field}

\author{V.~S.~Manko\,$^\dag$ and E.~Ruiz$\,^\ddag$}
\address{$^\dag$Departamento de F\'\i sica, Centro de Investigaci\'on y de
Estudios Avanzados del IPN, A.P. 14-740, 07000 M\'exico D.F.,
Mexico\\$^\ddag$Instituto Universitario de F\'{i}sica
Fundamental y Matem\'aticas, Universidad de Salamanca, 37008 Salamanca, Spain}

\begin{abstract}
The equilibrium configurations between a Schwarzschild black hole
and a hyperextreme Kerr object are shown to be described by a
three-parameter subfamily of the extended double-Kerr solution.
For this subfamily, its Ernst potential and corresponding metric
functions, we provide a physical representation which employs as
arbitrary parameters the individual Komar masses and relative
coordinate distance between the sources. The calculation of
horizon's local angular velocity induced in the Schwarzschild
black hole by the Kerr constituent yields a simple expression
inversely proportional to the square of the distance parameter.
\end{abstract}


\maketitle

%\twocolumn

\section{Introduction}

The general solution of the extended double--Kerr equilibrium problem is represented by the Ernst complex potential of the form \cite{MRu}
\bea \E&=&(\Lambda+\Gamma)/(\Lambda-\Gamma), \quad
\Lambda=\sum_{1\leq i<j\leq4}\lambda_{ij}r_ir_j, \quad
\Gamma=\sum_{i=1}^4\gamma_ir_i,\nonumber\\
\lambda_{ij}&=&(-1)^{i+j}(\a_i-\a_j)(\a_{i'}-\a_{j'})X_iX_j, \quad
(i',j'\neq i,j;\,\, i'<j'), \nonumber\\
\gamma_{i}&=&(-1)^{i}(\a_{i'}-\a_{j'})(\a_{i'}-\a_{k'})
(\a_{j'}-\a_{k'})X_i, \quad (i',j',k'\neq i;\,\, i'<j'<k'),
\nonumber\\ r_i&=&\sqrt{\rho^2+(z-\a_i)^2}, \label{E_pot} \eea
where the constants $\a_i$, $i=1,2,3,4$, can take on arbitrary real values or occur in complex conjugate pairs, and the quantities $X_i$ are defined as follows:
\bea
&&X_1=\frac{v_1-\phi}{\phi^{-1}-v_1}, \quad
X_2=\frac{1-\phi v_1}{\phi^{-1}v_1-1}, \quad X_3=\frac{1+i\phi v_4}{1-i\phi^{-1} v_4}, \quad
X_4=\frac{-\phi+i v_4}{\phi^{-1}+i v_4}, \nonumber\\
&&v_1=\epsilon_1\left[\frac{(\a_1-\a_3)(\a_1-\a_4)}
{(\a_2-\a_3)(\a_2-\a_4)}\right]^{1/2}, \quad v_4=\epsilon_4\left[\frac{(\a_1-\a_4)(\a_2-\a_4)}
{(\a_1-\a_3)(\a_2-\a_3)}\right]^{1/2},
\label{Xi} \eea
$\phi$ being an arbitrary complex constant of modulus one ($\phi\bar\phi=1$); $\epsilon_1=\pm1$ and $\epsilon_4=\pm1$.

The above potential $\E$ satisfies the Ernst equation \cite{Ern} and describes the equilibrium configurations of two arbitrary aligned Kerr sources which can be black holes, hyperextreme objects or their combinations. Because the four parameters $\a_i$ determine location of the sources on the symmetry axis, they can always be parametrized by three  arbitrary real constants, so that formulas (\ref{E_pot}) and (\ref{Xi}) involve, accounting for $\phi$, {\it four} real independent parameters which can be related to the masses and angular momenta of the Kerr constituents. In the paper \cite{MRu2} the Komar quantities \cite{Kom} were calculated for each component of an equilibrium configuration and the following general equilibrium law for two Kerr constituents was established:
\bea &&J+s\Bigl(\frac{j_1}{m_1}+\frac{j_2}{m_2}\Bigr)+\delta\epsilon(M+s)^2=0, \nonumber\\ &&M=m_1+m_2, \quad J=j_1+j_2, \quad \delta=\pm 1, \quad \epsilon=\pm 1, \label{blaw}  \eea
which indicates at which separation distance $s$ the equilibrium of spinning sources occurs for some given Komar masses $m_l$, $l=1,2$, and Komar angular momenta $j_l$.

The above equilibrium law raises an interesting question of whether balance is still possible when the angular momentum of one of the constituents, say $j_1$, is equal to zero? On the one hand, one is tempted to say that equilibrium in this case is impossible because the spin--spin repulsive force which ensures balance with the gravitational attractive force and emerges due to the interaction of sources' angular momenta must have zero value. On the other hand, as was already observed in \cite{Var}, in the non--equilibrium configurations composed of a rotating and a non--rotating black holes kept apart by a strut (this type of stationary subextreme systems is covered by the usual double--Kerr solution of Kramer and Neugebauer \cite{KNe}) the spinning black hole necessarily involves in rotation the horizon of the non--rotating black hole, so that the spin--spin repulsive force in such configurations is still present. But whether this `secondary' spin--spin interaction is sufficient for removing a strut?

These are some of the questions that will be answered in the
present paper. We will demonstrate that a Schwarzschild black hole
hole can freely levitate above the super--spinning Kerr object (in
spite of a recent claim \cite{Bon} on the non-existence of the
`Schwarzschild-Kerr' equilibrium configurations made on the basis
of a specific approximation scheme), and we will obtain the
general exact solution describing this specific equilibrium model.
In Sec.~II we reparametrize the `Schwarzschild--Kerr' equilibrium
problem by using the Komar masses of the constituents and the
separation distance as arbitrary real parameters, and derive a
representation of the Ernst complex potential and of all metric
functions in terms of these constant quantities. In Sec.~III the
results of Sec.~II are applied to the analysis of the physical
properties of the levitating Schwarzschild black hole. Concluding
remarks are given in Sec.~IV.

\section{The solution and metric functions}

The derivation of the solution describing equilibrium configurations of our interest partially simplifies if we take into consideration that

($i$) both Komar masses of our solution must have positive values: $m_1>0$, $m_2>0$;

($ii$) it is well known that in the double--Kerr solution the equilibrium states between two subextreme constituents with positive Komar masses do not exist \cite{DHo,MRu}, which means that the spinning partner of the Schwarzschild black hole in the binary system can only be a hyperextreme Kerr constituent whose location is defined by a complex conjugate pair $\a_4=\bar\a_3$ (see Fig.~1a);

($iii$) from (\ref{blaw}) it follows that, when $j_1=0$, the equilibrium condition can be solved most easily with respect to the remaining angular momentum $j_2$, hence the set of physical parameters of our solution is likely to be comprised by the individual Komar masses $m_1$ and $m_2$ jointly with the relative coordinate distance $s$ (three parameters in total).

With these remarks in mind, we now solve the equilibrium condition (\ref{blaw}) for $j_2$:
\be j_1=0 \quad \Longrightarrow \quad j_2=\epsilon\frac{m_2(s+m_1+m_2)^2}{s+m_2}, \quad \epsilon=\pm 1, \label{j2} \ee
where we have taken into account that, according to the paper \cite{MRu2}, we must choose $\delta=-1$ in (\ref{blaw}) for our concrete `subextreme--hyperextreme' configuration, the factor $\epsilon$ now defining the orientation of rotation.

The next step is to express the quantities $\a_i$ and $X_i$ in terms of $m_1$, $m_2$ and $s$. This can be done with the aid of the formulas for the Komar masses and angular momenta elaborated in the paper \cite{MRu2}. Then, after a very tedious but straightforward algebra we finally get the following concise expressions for $\a_i$ and $X_i$:
\bea \a_1&=&s+m_1, \quad \a_2=s-m_1, \quad \a_3=-i\sigma, \quad \a_4=+i\sigma, \nonumber\\ X_1&=&i\epsilon, \quad X_2=\frac{s+m_2+i\epsilon m_1}{m_1+i\epsilon(s+m_2)}, \nonumber\\ X_3&=&-\frac{s+m_1+m_2(1-i\epsilon)-\epsilon\s}{s+m_1+m_2(1+i\epsilon)+\epsilon\s}, \quad X_4=-\frac{s+m_1+m_2(1-i\epsilon)+\epsilon\s}{s+m_1+m_2(1+i\epsilon)-\epsilon\s}, \nonumber\\ \s&=&\sqrt{s^2-m_1^2+2s[m_1(1+\mu)+m_2]}, \quad \mu:=\frac{m_1}{s+m_2}. \label{Xi_rep} \eea
Note that for the derivation of (\ref{Xi_rep}) we have found it advantageous to place the hyperextreme Kerr constituent at the origin of coordinates, the Schwarzschild black hole locating above it (see Fig.~1b).

The substitution of (\ref{Xi_rep}) into (\ref{E_pot}) yields the desired representation of the Ernst potential determining the `Schwarzschild--Kerr' equilibrium configurations:
\bea \E&=&(A-B)/(A+B), \nonumber\\ A&=&m_2\sigma\mu[(R_++R_-)(r_--r_+)-2(R_+R_--r_+r_-)] +i\sigma^2(R_+-R_-)(r_++r_-)-\epsilon s(1+\mu) \nonumber\\ &\times&[m_2\mu(R_+-R_-)(r_--r_+)+i\epsilon m_2(R_+-R_-)(r_++r_-)+i\sigma(R_++R_-)(r_++r_-)], \nonumber\\ B&=&2\epsilon s\mu \{[m_2(s+m_1+m_2)(1-i\epsilon\mu)-i\epsilon\sigma^2](R_--R_+) -i\sigma[s+m_1+(1+i\epsilon)m_2] \nonumber\\ &\times&(R_++R_-)\} -2i\epsilon m_2 s\sigma[(1+\mu)^2r_++(1-i\epsilon\mu)^2r_-], \label{E_pot2} \eea
where we have introduced new notations for the functions $r_i$:
\be R_\pm=\sqrt{\rho^2+(z\pm i\sigma)^2}, \quad r_\pm=\sqrt{\rho^2+(z-s\pm m_1)^2}. \label{Rr} \ee

It is easy to check that if $m_2=0$, the potential (\ref{E_pot2}) reduces to that defining the Schwarzschild solution with the additional shift $s$ along the symmetry axis, and if $m_1=0$, it describes the hyperextreme Kerr solution \cite{Ker} with $j_2^2>m_2^4$. The axis value $e(z)$ of the potential obtained has the form ($z>s+m_1$)
\be e(z)=\frac{z^2-(1+i\epsilon)(s+m_1+m_2)z-(s+m_1)^2+(1+i\epsilon) (\sigma^2-m_1m_2-m_2s)}{z^2+(1-i\epsilon)(m_1+m_2-i\epsilon s)z+i\epsilon(s-m_1) [s+m_1+(1+i\epsilon)m_2]}. \label{axis} \ee

The calculation of the metrical fields $f$, $\gamma$ and $\omega$ entering the stationary axisymmetric line element
\be
\d s^2=f^{-1}[e^{2\gamma}(\d\rho^2+\d z^2)+\rho^2\d\varphi^2]-f(\d
t-\omega\d\varphi)^2 \label{Papa} \ee
can be performed in our case with the aid of the general formulas of the paper \cite{MRS}, yielding the following final result:
\bea f&=&\frac{A\bar A-B\bar B}{(A+B)(\bar A+\bar B)}, \quad e^{2\gamma}=\frac{A\bar A-B\bar B}{K_0R_+R_-r_+r_-}, \quad \omega=\omega_0-\frac{2{\rm Im}[G(\bar A+\bar B)]}{A\bar A-B\bar B}, \nonumber\\ G&=&-zB+4m_2 s\sigma\mu r_+r_-+s(1+\mu)^2r_-\{\sigma[m_2+(1-i\epsilon)s](R_++R_-) \nonumber\\ &-&[m_2(\epsilon s+im_1)+(\epsilon-i)s(s+m_1+m_2)](R_+-R_-)\} +is(1+i\epsilon\mu)r_+ \nonumber\\ &\times&\{[s(1+\mu)+m_2-\epsilon\sigma](s+m_1-i\sigma)R_+- [s(1+\mu)+m_2+\epsilon\sigma](s+m_1+i\sigma)R_-\} \nonumber\\ &-&2i\epsilon m_2s\sigma[(s+m_1)(1+\mu)^2r_++(s-m_1)(1-i\epsilon\mu)^2r_-] \nonumber\\ &-&2\epsilon s\mu\{[m_2(s+m_1+m_2)(1-i\epsilon\mu)-i\epsilon\sigma^2] [2s(R_+-R_-)+i\sigma(R_++R_-)] \nonumber\\ &+&i\sigma[s+m_1+(1+i\epsilon)m_2] [2s(R_++R_-)+i\sigma(R_+-R_-)]\}, \nonumber\\ K_0&=&16s^2\sigma^2(1+\mu)^2, \quad \omega_0=2\epsilon(s+m_1+m_2). \label{metric}  \eea

Formulas (\ref{metric}) define an asymptotically flat metric regular on all parts of the symmetry axis outside the location of sources, i.e., on its upper part ($\rho=0$, $z>s+m_1$), its lower part ($\rho=0$, $z<0$), and on the segment separating the sources ($\rho=0$, $0<z<s-m_1$), the metric functions $\gamma$ and $\omega$ on these parts of the axis taking zero values. Therefore, the Schwarzschild black hole and hyperextreme Kerr constituent indeed form an equilibrium configuration due to a specific realization of the spin--spin interaction mechanism which will be discussed in the next section.

\section{Physical properties of a levitating black hole}

We first note that the application of the solution derived in the previous section to concrete equilibrium configurations of a Schwarzschild black hole and a Kerr hyperextreme object is very simple: one only needs to choose some particular values of the masses $m_1$ and $m_2$, together with the value of the distance $s$ ($s>m_1$) at which equilibrium must occur, and then find from formula (\ref{j2}) the corresponding value of the angular momentum $j_2$ ensuring equilibrium. The space--time geometry of that particular configuration is determined by formulas (\ref{E_pot2}), (\ref{Rr}) and (\ref{metric}).

Turning now to the general properties of our solution, it should be emphasized that the levitating Schwarzschild black hole does not contribute to the general angular momentum of the system, and this fact can be readily verified, e.g., with the help of Tomimatsu's angular--momentum formula \cite{Tom}. At the same time, it is not difficult to calculate the horizon's local angular velocity $\Omega$ \cite{Car} induced by the Kerr constituent in the Schwarzschild black hole because this quantity is equal to the inverse value of the metric function $\omega$ evaluated on the horizon. From (\ref{E_pot2}), (\ref{Rr}) and (\ref{metric}) we get the following simple formula:
\be \Omega=\frac{\epsilon m_2}{2(s+m_2)(s+m_1+m_2)}, \label{Om} \ee
which means that $\Omega$ is inversely proportional to the square of the distance between the constituents and has the same sign as the angular momentum $j_2$. The corotating Kerr hyperextreme object and the Schwarzschild horizon give rise to the spin--spin repulsive force compensating the gravitational attraction of the constituents, which is in agreement with the general observations made in the aforementioned paper \cite{Var} by Varzugin. At the same time, an estimation made in \cite{Var} concerning $\Omega$ resulted in the induced angular velocity proportional to $s^{-3}$; however, such a more rapid, compared to our formula (\ref{Om}), decreasing of $\Omega$ with distance could be explained by the presence in \cite{Var} of a strut attached to the horizon and slowing down its velocity. Note that $\Omega$ of the black--hole constituent takes zero value (in addition to $j_1=0$ that we demanded from the very beginning) when $m_2=0$ (absence of the hyperextreme constituent) or $s=\infty$ (infinite separation), i.e., when the Schwarzschild horizon is not affected by frame--dragging.

For the calculation of the horizon area one can use the formula $4\pi m_1[-\omega_H^2\exp(2\gamma_H)]^{1/2}$, where $\omega_H$ and $\gamma_H$ are values of the functions $\omega$ and $\gamma$ on the horizon; after its application we get
\be A_H=16\pi m_1^2\Bigl(1+\frac{m_2}{s}\Bigr), \label{area} \ee
and the well--known expression for the area of the isolated Schwarzschild horizon is recovered from (\ref{area}) in the limit $s\to\infty$, or in the absence of the second body ($m_2=0$). Apparently, the entropy of the Schwarzschild black hole increases as a result of the interaction with the Kerr constituent.

It is worth noting that an induced angular velocity of the horizon is not the only peculiar feature the non--rotating black hole acquires due to the spin--spin interaction with a Kerr source. Thanks to the latter interaction, the levitating black hole also develops a stationary limit surface touching the event horizon at the points $\rho=0$, $z=s\pm m_1$. One may speculate in this respect that the usual Penrose process of the energy extraction from a rotating black hole \cite{Pen} could work in the case of our black--hole constituent too, most probably supplying it with a non--zero Komar angular momentum.

Since from the point of view of the Kerr geometry our planet
represents a hyperextreme spinning object, one in principle might
ask himself a question about whether any non--spinning mass could
freely float, say, above the Earth's north pole. Although,
intuitively, the answer to such question is `no' because of a
rather weak gravitational field at the Earth's surface, it seems
instructive to analyze the situation in more detail using the
balance condition (\ref{j2}). For this purpose we first put in
(\ref{j2}) $\epsilon=1$ and pass to the Komar quantities in CGS
units via the substitutions $m_{1,2}\to\tilde m_{1,2}G/c^2$,
$j_{2}\to\tilde j_{2}G/c^3$, $G$ being the gravitational constant
and $c$ the speed of light, thus making it possible to carry out
rough estimations by assigning to $\tilde m_{1,2}$ and $\tilde
j_{2}$ particular Newtonian values. Then, confining ourselves to
the simplest model in which the Earth is a rigidly rotating sphere
of uniform density, we can put $\tilde m_{2}=6\times 10^{27}$~g
and calculate the corresponding angular momentum $\tilde
j_{2}\simeq 7.166\times 10^{40}$~g~cm$^{2}$~s$^{-1}$. With these
values, and with the distance parameter $s$ approximately equal to
(but greater than) Earth's radius $6.4\times 10^{8}$~cm, we now
solve the condition (\ref{j2}) for $\tilde m_1$ and arrive at a
negative value, which means that physically relevant
`Earth--static mass' equilibrium configurations involving a
levitating constituent are not possible.

At the same time, imagine a hyperextreme Kerr object characterized
by the above Earth's values $\tilde m_2$, $\tilde j_2$ and the
size of the order of Earth's Schwarzschild radius ($<1$ cm). In
this case, for fixed $\tilde m_2$ and $\tilde j_2$, we can assign
different positive values to $\tilde m_1$ and find from (\ref{j2})
the corresponding values of $s$ defining the equilibrium states;
of course, instead of assigning values to $\tilde m_1$, if
necessary, one can also vary $s$ and find the corresponding values
of $\tilde m_1$ by solving (\ref{j2}). A simple analysis then
shows that, quite surprisingly, although no equilibrium
configurations exist for any positive $\tilde m_1$ at
$s>397.69$~cm, all equilibrium states for $0<\tilde
m_1<2.249\times10^{21}$~g occur within a very narrow interval
$397.686533\le s<397.69$~cm, while the equilibrium states of
non--rotating black holes with $2.249\times10^{21}<\tilde
m_1<4.63\times10^{27}$~g are all covered by the interval $397\le
s<397.686533$~cm! This clearly demonstrates how strong the
spin--spin repulsive force can be and how rapidly it can grow with
diminishing separation distance. The fact that in the particular
equilibrium configurations considered above the values of $s$ by
far exceed the Schwarzschild radii of the constituents suggests
likeliness of the formation of analogous configurations in the
vicinities of compact astrophysical objects such as, for instance,
neutron stars \cite{STe}. However, the observational aspects of
equilibrium states need further investigation.

\section{Conclusions}

In the present paper we have demonstrated a remarkable property of the Schwarzschild black hole to form equilibrium configurations with a hyperextreme Kerr source. Our analysis is essentially based on the three--parameter subfamily of the extended double--Kerr solution for which we have been able to work out a simple parametrization involving Komar masses and separation distance as arbitrary parameters. Although the Komar angular momentum of the Schwarzschild black--hole constituent remains equal to zero all the time, the black hole horizon becomes involved in rotation due to the stationarity of the spacetime, and the resulting spin--spin interaction with the Kerr source turns out sufficient for counteracting the gravitational attractive force and attaining equilibrium of the two constituents.

In view of the presence in our solution of a naked singularity which is a well--known characteristic of the hyperextreme Kerr object, it would be interesting to see whether the five--dimensional black rings \cite{ERe,ERe2} could replace the hyperextreme Kerr constituent in the 4+1--analogs of our equilibrium configurations as the black rings can have arbitrarily large angular momenta for fixed masses but at the same time they do not develop naked singularities. It seems that having a slight generalization of the black Saturn solution \cite{EFi} in which the black hole and the black ring could be located in two different planes would probably be enough for reproducing our main results obtained for the levitating Schwarzschild black hole. The construction of such generalization looks feasible if one takes into account a formal similarity of the solution--generating techniques which are being currently used in the four- and five--dimensional gravities \cite{TIM}.

\section*{Acknowledgments}

We would like to thank Professor M.A. Shamsutdinov for a useful suggestion, and Professor~W.G.~Unruh for valuable comments on the earlier version of this paper. Our work was supported by Project FIS2006-05319 from Ministerio de Ciencia y Tecnolog\'\i a, Spain, and by the Junta de Castilla y Le\'on under the ``Programa de Financiaci\'on de la Actividad Investigadora del Grupo de Excelencia GR-234'', Spain.

\newpage

\begin{figure}[htb]
\centerline{\epsfysize=80mm\epsffile{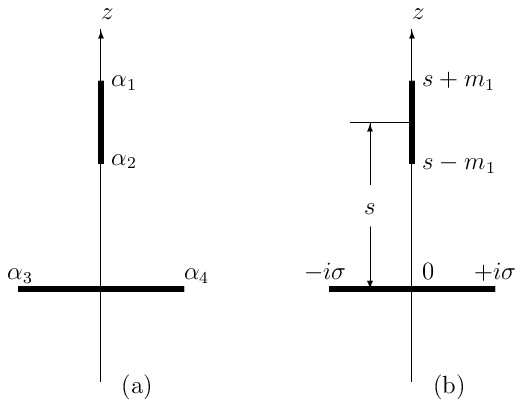}} \caption{Location of the Schwarzschild black hole (a bar on the symmetry axis) and Kerr hyperextreme object (a cut perpendicular to the axis) parametrized by two different parameter sets: (a) using the canonical parameters $\a_i$, and (b) using a physical set of parameters, with $\sigma$ defined by formula (\ref{Xi_rep}).}
\end{figure}

\end{document}